\documentstyle[prl,aps,floats,psfig,twocolumn]{revtex}

\begin{document}

\draft

\title{Test of Parity-Conserving Time-Reversal Invariance \\ Using
        Polarized Neutrons and Nuclear Spin Aligned Holmium\footnote{This 
        work was supported
        in part by the U.S. Department of Energy, Office of High Energy and
        Nuclear Physics, under contracts DE-FG09-88-ER40411 and
        DE-FG05-91-ER40619.}}
\author{P.\,R. Huffman\footnote{Present address: Department of Physics,
        Harvard University, Cambridge, MA 02138}, N.\,R. Roberson, and
        W.\,S. Wilburn}
\address{Physics Department, Duke University, Durham, NC 27708-0305 \\
        and Triangle Universities Nuclear Laboratory, Durham, NC 27708-0308}
\author{C.\,R. Gould, D.\,G. Haase, C.\,D. Keith\footnote{Present address:
        IUCF, 2401 Milo B. Sampson Lane, Bloomington, IN 47405},
        B.\,W. Raichle, M.\,L. Seely, and J.\,R. Walston}
\address{Physics Department, North Carolina State University, 
        Raleigh, NC 27695-8202 \\
        and Triangle Universities Nuclear Laboratory, Durham, NC 27708-0308}
\date{\today}

\maketitle

\begin{abstract}
A test of parity-conserving, time-reversal non-invariance (PC TRNI) has been 
performed  in 5.9~MeV polarized neutron transmission through nuclear spin 
aligned holmium.  The experiment searches for the T-violating five-fold
correlation via a double modulation technique --- flipping the
neutron spin while rotating the alignment axis of the holmium.
Relative cross sections for spin-up and spin-down neutrons are 
found to be equal to within $1.2 \times 10^{-5}$ (80\% confidence). 
This is a two order of magnitude improvement compared to traditional detailed 
balance studies of time reversal, and represents the most precise 
test of PC TRNI in a dynamical process.  
\end{abstract}


Parity-conserving, time-reversal non-invariance (PC TRNI) arises only through 
second-order weak effects within the Standard Model.  As such, observables 
from these interactions are expected to be extremely small \cite{Her88}.  
Nevertheless, experimental bounds are much less stringent.  Previously, the 
most precise dynamical bound came from the detailed balance studies of the 
reaction $^{24}\mbox{Mg}(\alpha, p)^{27}\!\mbox{Al}$ and its inverse 
\cite{Bla83}, where relative differential cross sections were found to be 
equal to within $5.1 \times 10^{-3}$ (80\% confidence).
  
In this letter, we present results from an improved search for PC TRNI, 
using polarized neutron transmission through a rotating, cryogenically 
aligned, $^{165}$Ho target \cite{Koster}.  The measurement tests reciprocity, 
or more colloquially, ``running the movie backwards''. If reciprocity holds, 
the total cross sections  will be equal for vertically polarized  spin-up 
and spin-down neutrons transmitted through a tensor polarized target 
whose alignment axis lies in the horizontal plane at 45 degrees with 
respect to the beam direction \cite{Kabir}. We find the relative cross 
sections for 5.9~MeV neutrons to be equal to within $1.2 \times 10^{-5}$ 
(80\%), which, compared to detailed balance studies, is a factor of four 
hundred improvement in a measurement of PC TRNI relative cross sections.

A more fundamental comparison of results of different experiments at low 
energies is made in terms of a meson-exchange model.  Simonius has shown 
that parity-conserving, time-reversal violation arises only through charged 
meson exchanges, with the major contribution arising from the $\rho$ meson 
\cite{Sim75}. We take advantage of a recently developed model incorporating 
T-violating $\rho$ meson exchange \cite{Eng94} to interpret our result. 

The present measurement consists of a search for the five-fold correlation 
(FC) term $\vec{s}\cdot(\vec{I}\times\vec{k})(\vec{I}\cdot\vec{k})$ in the 
neutron-nucleus forward scattering amplitude. Here, $\vec{s}$\/ is the spin of 
the neutron, $\vec{k}$ is the momentum of the neutron, and $\vec{I}$ is the 
spin of the holmium target.  The total cross section for neutrons polarized 
parallel/anti-parallel ($+$/$-$) to the direction $\vec{I} \times \vec{k}$ 
is \cite{Hni94b}
\begin{eqnarray}
\sigma^{\pm}_{T}(\theta) & = & \sigma_{0} (1 + \tilde{t}_{20}(I)\,\sigma_{2}\,
                            P_{2}(\cos\theta) \\ \nonumber
                         &   & \quad {} \pm \tilde{t}_{10}(s)\, 
                         \tilde{t}_{20}(I)\,A_{5}\,\sin2\theta),
\end{eqnarray}
where $\tilde{t}_{10}(s)$ is the polarization of the neutron beam, 
$\tilde{t}_{20}(I)$ is the tensor alignment of the holmium target with 
respect to its crystal symmetry axis, $\sigma_{0}$ is the unpolarized cross 
section, $\sigma_{2}$ is the deformation effect cross section, $A_{5}$ is the
PC TRNI spin-correlation coefficient, and $\theta$ is the angle between
the alignment axis of the holmium crystal and the beam direction.  

The FC term is isolated by reversing the spin of the neutron beam, and 
simultaneously rotating the holmium alignment axis.  A sequence of 
measurements of the transmission asymmetry
\begin{equation}
\label{eqn:asymmetry}
\epsilon_5(\theta) = \frac{N^{+}(\theta) - N^{-}(\theta)}{N^{+}(\theta) +
N^{-}(\theta)}
\end{equation}
($N^{\pm}(\theta) = N_{0} e^{-n\sigma_{T}^{\pm}(\theta)}$, where $n$ 
is the target thickness and $N_{0}$ is the incident flux) is fit to the 
form $a_{0} + a_{2}\,\sin 2\theta$ ($a_{0}$, $a_{2}$ constants) to extract 
the spin-correlation coefficient
\begin{equation}
\label{eqn:A5}
A_{5} = \frac{a_{2}(1+\phi)}
             {\tilde{t}_{10}(s)\,\tilde{t}_{20}(I)\,n\,\sigma_{0}},
\end{equation} 
where $n=0.065$~atoms/b, and $\sigma_{0}$ is the unpolarized cross section 
(5.1~b at $E=5.9$~MeV). The factor $\phi$ accounts for the small number of 
gamma rays detected and counted as neutrons (see later). 

Systematic effects associated with time drifts, spin and beam 
misalignments, and finite geometry effects are analyzed in detail 
elsewhere \cite{Huffman}.  In general, none lead to a $\sin2\theta$ 
modulation at the level investigated here.  In particular, sequential 
interactions which mimic the FC signal exactly are second order in 
the weak interaction, and are negligible to the accuracy of the present 
experiment.

\begin{figure}[htbp]
\centerline{\psfig{figure=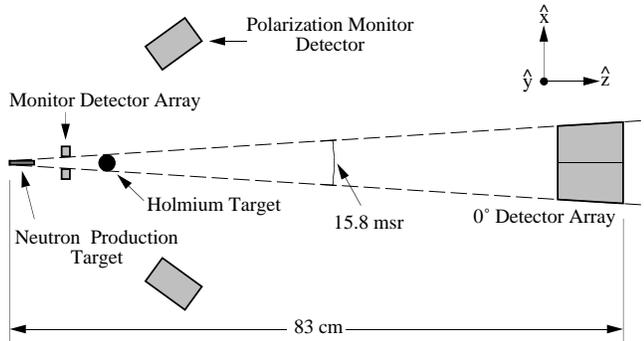,width=3.375in,silent=}}
\caption{The experimental setup for the time-reversal measurement.
A vertically ($\hat{y}$) polarized neutron beam with 
momentum $\hat{k}$, directed along $\hat{z}$, is produced via the 
$^{2}\mbox{H}(\vec{d},\vec{n})^{3}\mbox{He}$ reaction, transmitted 
through the aligned holmium target, and detected at $0^{\circ}$.  The 
dashed lines depict the solid angle subtended by the neutron 
detectors. All components and distances are drawn to scale.}
\label{fig:setup}
\end{figure}

A schematic of the experimental arrangement is shown in 
Figure~\ref{fig:setup}. The count rate and accuracy improvement compared to our 
previous work \cite{Koster} result from: a) using the
$^{2}\mbox{H}(\vec{d},\vec{n})^{3}\mbox{He}$ reaction instead of the 
$^{3}\mbox{H}(\vec{p},\vec{n})^{3}\mbox{He}$ reaction, b) using a 
cryogenically cooled deuterium gas cell, c) moving source and 
detectors into a close, unshielded geometry, and d) longer run times.

The cooled deuterium gas cell is  a 0.851~cm diameter by 3.18~cm long 
cylinder filled with 8~atm of deuterium gas.  The vector and tensor polarized 
deuteron beam from the TUNL tandem accelerator enters the cell through a 
15.2~$\mu$m thick Havar window and is stopped in a 0.051~cm sleeve of 
gold surrounding the cell walls.  The cell is cooled to 168~K through a 
2.54~cm diameter copper cold finger extending into a liquid nitrogen bath.  The 
cell temperature is stabilized by a feedback heater to $\pm 0.5$~K to 
minimize gas density fluctuations due to beam heating.  A 2.0~$\mu$A 
beam of 4.9~MeV deuterons produced $\sim 10^{5}$~neutrons/cm$^{2}\cdot$s 
at the front surface of the $0^{\circ}$ detector array with a mean energy 
of 5.9~MeV and a spread of 2.6~MeV\@.  A pair of liquid scintillator 
detectors are located at $\pm 36^{\circ}$ with respect to the beam 
direction to monitor the neutron polarization via the left-right analyzing 
power of the neutron production reaction. 

Holmium is chosen as a target material because it is monoisotopic and can be 
cryogenically aligned in the absence of an external magnetic field.  
The target consists of a cylindrical single-crystal sample (2.3~cm
diameter, 2.8~cm in height) with its {\it c}-axis (or alignment axis)
oriented perpendicular to the cylinder axis. The crystal is mounted with its
cylinder axis along $\hat{y}$, placing its alignment axis in the $x-z$ plane. 
The target is rotated about the cylinder axis thereby defining the angle 
$\theta$ between the alignment axis and the beam direction.  The sample is 
cooled to $\sim 150$~mK using a $^{3}$He -- $^{4}$He dilution refrigerator, 
causing spontaneous nuclear alignment (90\%) due to the interaction of the 
magnetic dipole and electric quadrupole moments with the unpaired electrons 
through the large internal hyperfine field. The vector polarization of 
the sample due to the earth's magnetic field is $10^{-4}$, and is 
negligible for the present measurement.  The alignment of the holmium was 
measured using thermometry and verified with independent  measurements of the 
deformation effect cross section \cite{Kos94}.  These measurements 
were performed at 9.5~MeV where the deformation effect cross section is large 
($\sim -500$~mb).  The time-reversal measurements were carried out 
at 5.9~MeV where the deformation effect is small ($< 5$~mb), but 
sensitivity to TRNI is a maximum \cite{Eng94}.

The intense neutron fluxes at both the monitor and $0^{\circ}$ detectors 
required the development of two four-detector arrays of plastic scintillator 
neutron detectors.  The monitor detector array is 1.27~cm thick and detects 
neutrons in a ``halo'' surrounding the solid angle subtended by the 
$0^{\circ}$ detector array.  The $0^{\circ}$ array is $12.7 \times 12.7 
\times 10.2$~cm thick and divided into four equal segments.  Detectors are 
operated in pulse mode.  Since plastic scintillators are used, pulse shape 
discrimination between neutrons and gamma rays is not possible.  Thus, the 
ratio of gamma rays to neutrons is determined independently using 
time-of-flight techniques, and is measured to be $\phi = 0.064 \pm 0.001$. 

The data for the time-reversal measurements are collected using a double 
modulation technique, where the neutron spin direction is reversed every 
100~ms in the eight-step sequence ${}+{}-{}-{}+{}-{}+{}+{}-{}$.  After 256 
eight-step sequences, the target alignment axis is rotated to a new angle in 
the sequence $-180^{\circ}$ to $+180^{\circ}$ and back to $-180^{\circ}$ in 
$22.5^{\circ}$ steps.  Both the spin-flip and target rotation sequences are 
chosen to minimize time-dependent drifts that can arise in the transmission 
asymmetry.  

The neutron transmission data are corrected for dead-time and cross talk 
between detector pairs before normalizing to the counts in the monitor 
detector array.  An asymmetry is formed (Eq.~\ref{eqn:asymmetry}) for each 
eight-step sequence and averaged over the entire set of 256 eight-step 
sequences.  A constant background ($\sim 5 \times 10^{-4}$) term  arises,
due to the difference in tensor polarizations of the deuteron beam between 
the two polarization states, which results in different neutron fluxes.
The T-violating FC term, if it exists,  appears in the data as a 
$\sin2\theta$ oscillation on top of this constant background. It is 
this angle modulation, in combination with the rapid neutron spin flip, 
that makes possible the high precision achieved in this experiment. 

\begin{figure}[ht]
\centerline{\psfig{figure=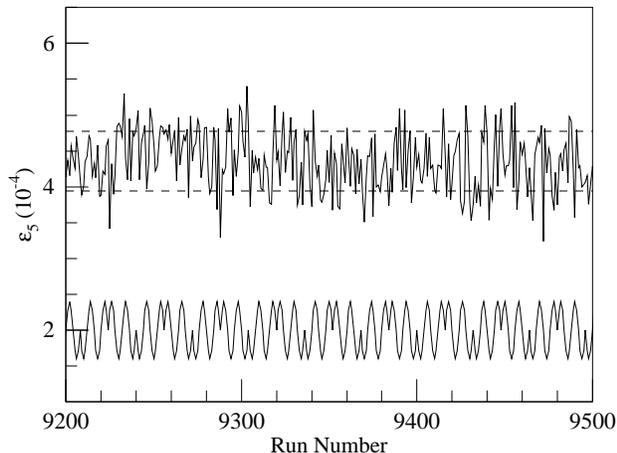,silent=}}
\caption{Monitor normalized detector asymmetry $\epsilon_{5}$ 
as a function of run number for a 
300 run subset of data.  Each run corresponds to four minutes of data at 
a given angle, taken in the sequence $-180^{\circ}\rightarrow +180^{\circ} 
\rightarrow -180^{\circ}$ in $22.5^{\circ}$ steps.  The dashed lines 
indicate the $\pm 1\sigma$ errors on the data.  The $\sin2\theta$ angle 
dependence associated with time-reversal violation is shown below the 
data with an arbitrary amplitude and a known 
phase.  Based on fits to these and the other data, we find the 
amplitude of the time-reversal violating $\sin2\theta$ term to be $1.1 \pm 
1.0 \times 10^{-6}$.}
\label{fig:ay}
\end{figure}

The data were accumulated over a period of one week, and correspond
to a total of $\sim 10^{12}$~neutrons detected.  The monitor normalized 
asymmetry $\epsilon_{5}(\theta)$ for a sequence of 600 runs is shown in 
Figure~\ref{fig:ay}.  Each run corresponds to 256 eight-step sequences at 
a given angle and the subset shown represents 25\% of the data.  The 
asymmetry is fit to the form $a_{0}+a_{2}\sin2\theta$ using least squares.  
A value of $a_{2} = (1.1 \pm 1.0) \times 10^{-6}$ is extracted from the 
data.  The chi-square per degree of freedom is 0.9994, indicating no 
significant random errors other than those associated with counting 
statistics.  Using the 
measured polarizations ($\tilde{t}_{10}(s) = 0.67 \pm 0.05$, 
$\tilde{t}_{20}(I) = 0.62 \pm 0.05$) and the measured ratio of gamma rays 
to neutrons ($\phi = 0.064 \pm 0.001$), the asymmetry is converted to the 
spin-correlation coefficient $A_{5}$ (Eq.~\ref{eqn:A5}), yielding
\begin{equation}
A_{5} = (8.6 \pm 7.7) \times 10^{-6},
\end{equation}
consistent with time-reversal invariance. At  $45^{\circ}$, the relative 
spin-up versus spin-down cross section difference is less than 
$1.2 \times 10^{-5} $ (80\% confidence).

The most natural parameter for describing PC TRNI is $\bar{g}_{\rho}$, the 
ratio of T-violating to T-conserving coupling constants for $\rho$ exchange.  
Using a recent theoretical analysis by Engel {\it et al.}, $A_{5}$ can be 
directly converted into $\bar{g}_{\rho}$ \cite{Eng94}.  In this analysis, 
the Simonius potential is used in a folding model calculation to generate an 
optical potential for $^{165}$Ho.  This optical potential is then used in a 
coupled-channels calculation to extract $A_{5}$ as a function of 
$\bar{g}_{\rho}$.  Using our value of $A_{5}$, we obtain $\bar{g}_{\rho} = 
(2.3 \pm 2.1) \times 10^{-2}$.  The FC term arises only from the valence 
proton in holmium, and therefore the aligned target PC TRNI experiments suffer 
from a 1/A suppression compared to one body nuclear effects such as parity
violation. Nevertheless, we see that an experiment with MeV neutrons is able 
to probe TRNI in the $\rho$-exchange part of the N-N potential at the level 
of a few per cent. 

Recent theoretical analyses \cite{Eng94,Fre87,Hax94,Bey93a} now allow 
comparison between TRI tests in various systems.  While the parameter 
$\bar{g}_{\rho}$ is the most natural point of comparison, an alternative 
quantity $\alpha_{T}$, the ratio of T-violating to T-conserving nuclear 
matrix elements has also been widely used in the past. The quantity 
$\alpha_{T}$ was introduced by the Rochester group, and bounds of order 
$10^{-3}$ were deduced by them from level spacing data in heavy nuclei
\cite{frenchprl}. It was considered initially that $\alpha_{T}$ was also 
an approximate measure of the relative strength of the PC TRNI in the N-N 
system. But more recent work has shown that  $\alpha_{T}$ and $\bar{g}_{\rho}$ 
are numerically quite different, and typically  
$\alpha_{T} = 0.012\bar{g}_{\rho}$ \cite{Hax94,Bey93a}.   

Previously, the detailed balance measurement of Blanke {\it et al.} 
\cite{Bla83} had provided the most stringent dynamical test of 
parity-conserving, time-reversal 
non-invariance.  In their experiment the cross sections for the reactions 
$^{24}\mbox{Mg}(\alpha, p)^{27}\!\mbox{Al}$ and its inverse were 
found to be equal to an accuracy of $\Delta=5.1 \times 10^{-3}$ 
(80\% confidence).  The experimental observable is not as simply related 
to the underlying nucleon-nucleon TRNI as in the FC experiment, and numerous 
statistical analyses \cite{Fre87,Boo86a,Har90,Bun93} have been performed over 
the years to extract more fundamental TRNI parameters from the data, such
as $\alpha_{T}$. The analyses of Refs.\cite{Fre87,Boo86a} give similar 
bounds, $\alpha_{T} \leq 3.5 \times 10^{-3}$ (95\%). Comparing to our results 
for $\bar{g}_{\rho}$ at the same confidence limit, we find the FC experiment 
therefore represents a factor of five improvement in a bound on 
$\bar{g}_{\rho}$.

While the present measurement represents the most precise test of PC TRNI to 
date in a dynamical system, other tests can in some cases provide more 
restrictive bounds indirectly. The most important of these other bounds 
comes from measurement of the atomic electric dipole moment of $^{199}$Hg 
($d \lesssim 1.3 \times 10^{-27}~e\cdot\mbox{cm}$ \cite{Jac93}). The 
observable in that case is both P- and T-violating, but sets constraints  
on the assumed PC TRNI interaction via weak corrections.  In a recent work, 
Haxton {\it et al.} \cite{Hax94} obtained a limit $\bar{g}_{\rho} \lesssim 
1 \times 10^{-2}$ at the 95\% confidence level from the $^{199}$Hg result.

Direct tests of PC TRNI continue to be discussed, both at high 
energies (0.5 GeV/c) in few-nucleon systems, and at low energies in resonance 
reactions in medium mass and heavy nuclei.  Measurements of the FC in 
$\vec{p} - \vec{d}$ scattering have been proposed for the new storage ring
facility COSY at Julich \cite{Evers}.  Few-nucleon systems allow 
for a clean theoretical interpretation, and first estimates indicate 
potentially an order of magnitude improvement in a direct test of PC TRNI 
\cite{Be93}.  Resonance tests hold the most promise for large enhancements, 
but are also theoretically the hardest to interpret. Both neutron and charged 
particle tests have been discussed \cite{TRI94}. An important consideration in 
such tests is the need to obtain data on more than just one resonance.  This 
is a realistic goal for the FC test in holmium, and for traditional detailed 
balance tests in charged particle reactions.  It continues to be an issue 
for P-violating TRNI neutron transmission tests, where to date only single 
isolated resonances have been identified in polarizable nuclei (for example, 
in $^{139}$La).

In summary, we have tested reciprocity in nuclear reactions by measuring a 
T-violating spin-correlation coefficient $A_5$ of $(8.6 \pm 7.7) \times 
10^{-6}$ in polarized neutron transmission through nuclear spin aligned 
holmium.  The measurement corresponds to a bound on a T-violating meson 
coupling of $\bar{g}_{\rho} \leq 5.8 \times 10^{-2}$ (95\%). This represents 
the most precise test of parity-conserving, time-reversal non-invariance in a 
dynamical process.

\end{document}